\documentclass[10pt,conference]{IEEEtran}
\IEEEoverridecommandlockouts
\usepackage{amsthm}
\usepackage{amsmath}
\usepackage{amssymb}
\usepackage{bm}
\usepackage{booktabs}
\usepackage{cite}
\usepackage{graphics, framed}
\usepackage{graphicx}
\usepackage{epsfig}
\usepackage{epstopdf}
\usepackage{subfigure}
\usepackage{url}
\usepackage{multimedia}
\usepackage{paralist}
\usepackage[printonlyused]{acronym}
\usepackage{units}
\usepackage{balance}
\usepackage{multirow}
\usepackage{color,soul}
\usepackage[ruled,vlined]{algorithm2e}
\usepackage{dsfont}

\usepackage[top=0.98in, bottom=0.98in, left=0.673in, right=0.673in]{geometry}

\newcommand{\FGR}[1]{Fig.~\ref{#1}}

\newcommand{\SEC}[1]{Section~\ref{#1}}

\renewcommand{\baselinestretch}{0.88}
\usepackage{xcolor}

\newtheorem{proposition}{Proposition}

\begin{document}

\title{Kolmogorov--Arnold Lookup Table Demappers for Amplitude Phase-Shift Keying over Nonlinear Satellite Links}
\author{\IEEEauthorblockN{K{\"{u}}r{\c{s}}at~Tekb{\i}y{\i}k\IEEEauthorrefmark{1}\IEEEauthorrefmark{2},  G{\"{u}}ne{\c{s}}~Karabulut~Kurt\IEEEauthorrefmark{2}, and Antoine~Lesage-Landry\IEEEauthorrefmark{1}\IEEEauthorrefmark{2}}
\IEEEauthorblockA{\IEEEauthorrefmark{1}GERAD \& Mila, Montr\'eal, Canada\\ \IEEEauthorrefmark{2}Poly-Grames Research Centre, Department of Electrical Engineering, Polytechnique Montr\'eal, Montr\'eal, Canada\\ Emails: \texttt{\{kursat.tekbiyik, gunes.kurt, antoine.lesage-landry\}@polymtl.ca}}}
\maketitle

\begin{abstract}

Next-generation high-throughput satellites increasingly demodulate and decode on board, so the per-symbol cost of the receiver constrains payload power and throughput. Soft demappers for nonlinear satellite links must therefore produce accurate bit log-likelihood ratios (LLRs) at low cost. We introduce a KAN-LUT demapper, a two-layer Kolmogorov--Arnold network (KAN) whose learned univariate edge functions are stored as 128-entry lookup tables (LUTs), so that inference needs only table reads, linear interpolation, and additions. We bound the LLR error introduced by the tables and evaluate the demapper on amplitude phase-shift keying (APSK) links with low-density parity-check (LDPC) coding and a nonlinear traveling-wave-tube amplifier. For 16-, 32-, and 64-APSK, KAN-LUT stays within 0.2~dB of a per-cluster Gaussian log-MAP (LogMAP-CG) demapper, while its cost relative to LogMAP-CG falls from 87\% to 27\% as the order grows. It matches or slightly exceeds multilayer perceptrons (MLPs) at about one fifth of their cost, remains competitive under operating-point mismatch, and needs only tens of kilobytes of tables. KAN-LUT is therefore a low-cost candidate demapper for regenerative payloads.
\end{abstract}

\begin{IEEEkeywords}
APSK, satellite communications, soft demapping, Kolmogorov--Arnold networks, nonlinear channels.
\end{IEEEkeywords}

\section{Introduction}
Satellite communication systems are moving from bent-pipe transponders toward very high-throughput, multi-beam, and regenerative architectures that serve broadband, backhaul, and non-terrestrial network traffic~\cite{Kodheli2021,BenYahia2025}. Their downlinks are power-limited, so the on-board high-power amplifier is operated close to saturation to maximize radiated power and efficiency. Signals with a high peak-to-average power ratio (PAPR), such as square quadrature amplitude modulation (QAM), then require a large input back-off (IBO) or suffer severe nonlinear distortion~\cite{DeGaudenzi2006}. For this reason, the DVB-S2 and DVB-S2X standards adopt amplitude and phase-shift keying (APSK)~\cite{ETSI2014,ETSI2024}. APSK is used in satellite links because its ring structure accommodates power-amplifier constraints while retaining spectral efficiency. With few rings and many points per ring, APSK has a lower PAPR than QAM of the same order and tolerates operation nearer saturation~\cite{DeGaudenzi2006}. A
traveling-wave-tube amplifier (TWTA) introduces amplitude and phase distortion; consequently, a receiver matched to an additive-noise channel can misestimate bit
likelihoods. A receiver must therefore account for the
compressed, non-Gaussian symbol clusters at the demapper input.

The per-cluster Gaussian log-MAP (LogMAP-CG) demapper uses fitted per-symbol Gaussian clusters and log-sum-exp bit metrics~\cite{Robertson1995}. MaxLog replaces log-sum-exp by maxima, reducing computation at a
worse operating point. Learned alternatives include QAM log-likelihood ratios (LLRs)~\cite{Shental2019}, impairment-aware deep neural networks~\cite{DobreLampe2022}, optical-channel metrics
\cite{Schaedler2020}, and joint QAM/APSK learning~\cite{Gansekoele2025}. A recent low-Earth-orbit receiver refines MaxLog centroids by K-means clustering and exploits constellation symmetry to cut 32-APSK demapping cost~\cite{Wu2026}.

A Kolmogorov--Arnold network (KAN) places learnable univariate functions on edges~\cite{Liu2024}, so each trained edge can be represented by a one-dimensional lookup table (LUT) such LUT evaluations have been used for low-latency field-programmable gate array (FPGA) inference~\cite{Hoang2026}, and KANs have been applied to nonlinear optical equalization~\cite{Fischer2025}. Inference then uses table reads, linear interpolation, and additions rather than dense multiply-accumulate layers. The KAN architecture can also mimic physical models~\cite{Tekbiyik2026}. Its storage grows linearly with the number of edges, whereas a direct LUT over
the in-phase, quadrature, and downlink signal-to-noise ratio (SNR) inputs grows exponentially with input dimension. To take advantage of this, we propose the KAN-LUT demapper, which approaches LogMAP-CG accuracy at lower cost than a multilayer perceptron (MLP).

\begin{figure*}[!t]
\centering
\includegraphics[width=\linewidth]{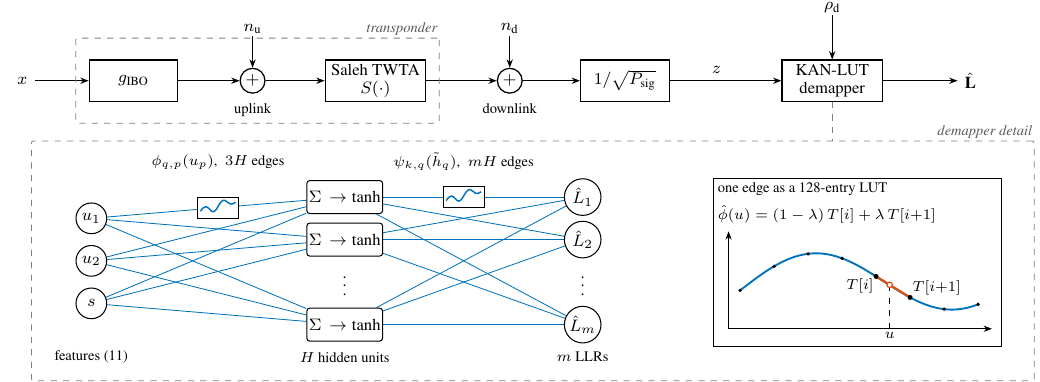}
\caption{Link model (top) and two-layer KAN-LUT demapper (bottom). Each univariate edge is a cubic spline stored as a 128-entry table and evaluated by linear interpolation (inset).}
\label{fig:system}
\end{figure*}

Reduced-complexity APSK demapping has mainly been studied for specific ring structures, such as the \(4+12\) and \(4+12+16\) constellations~\cite{LeeYoon2013,LeeJangYoon2012}. On the transmitter side, the TWTA distortion described by the Saleh model~\cite{Saleh1981} is usually compensated by predistortion or by shaping the constellation, increasingly with neural networks~\cite{Zayani2008,Kopyto2024,Kopyto2025}. At the receiver, the coded link is a bit-interleaved coded modulation system, and the generalized mutual information (GMI) of the demapper output measures how much information its LLRs pass to the decoder~\cite{Caire1998,Martinez2009}.

Receiver cost matters most when demodulation moves on board. Regenerative payloads demodulate and decode traffic in orbit, and next-generation designs target terabit-per-second throughput per satellite under tight power budgets~\cite{BenYahia2025}. The demapper runs once per received symbol, so its cost grows with the symbol rate. Carrying 1~Tbit/s with 32-APSK and the rate-3/4 short code needs about \(2.7\times10^{11}\) symbols/s. With the weighted costs in \SEC{sec:results}, LogMAP-CG would then need about \(2.8\times10^{14}\) operations per second, while KAN-LUT requires about half of that.

For such payloads, existing demappers leave a gap. Model-based demappers are accurate but expensive, while learned demappers are flexible but usually rely on dense layers whose cost is rarely compared with the classical baselines under the same accounting. We design a KAN whose edges are stored as LUTs to close this gap on a coded, nonlinear APSK link, with a view toward on-board demodulation in high-throughput regenerative satellites. We compare it with MLPs and classical demappers on equal terms~\cite{Yu2024}: the same decoder and test frames, an explicit operation count for every method, and paired statistical tests over seeds. Our contributions are given as follows:
\begin{enumerate}[{C}1]
\item We compare the complexity and frame error rate (FER) of KAN-LUT, MLP, LogMAP-CG, and MaxLog demappers on a nonlinear 32-APSK link and show that KAN-LUT approaches LogMAP-CG accuracy at about half its weighted cost and below a fifth of the cost of the best MLP, which reduces the on-board processing load.
\item We repeat the comparison for 16- and 64-APSK with independent model and operating-point selection and show that the cost advantage over LogMAP-CG grows with constellation order, the regime targeted by high-throughput DVB-S2X modes.
\item We evaluate GMI under IBO and uplink/downlink mismatch with Holm-corrected paired tests and show that a demapper trained at one operating point remains usable when amplifier back-off and link conditions drift in orbit.
\item We compare the memory of edge-wise KAN tables with a direct multidimensional LUT and show that the demapper fits in tens of kilobytes of on-chip memory instead of tens of megabytes.
\end{enumerate}

\section{System Model}
\subsection{APSK and Nonlinear Channel}
\FGR{fig:system} summarizes the link and demapper. Each unit-energy symbol \(x\) is scaled to the requested input back-off, perturbed by uplink noise, passed through
a memoryless Saleh TWTA~\cite{Saleh1981}, and corrupted by downlink noise. The Saleh amplitude-to-amplitude (AM/AM) and amplitude-to-phase (AM/PM) conversion is

{\allowdisplaybreaks
\begin{align}
g_{\mathrm{IBO}}&=\beta_{\mathrm a}^{-1/2}10^{-\mathrm{IBO}/20},\\
S(v)&=\frac{\alpha_{\mathrm a}|v|}{1+\beta_{\mathrm a}|v|^2}
\exp\Bigl(j\Bigl[\arg(v)+\frac{\alpha_\phi|v|^2}{1+\beta_\phi|v|^2}\Bigr]\Bigr), \label{eq:saleh}\\
P_{\mathrm{sig}}&=\frac{1}{|\mathcal X|}\sum_{x\in\mathcal X}|S(g_{\mathrm{IBO}}x)|^2,\\
z&=\frac{S(g_{\mathrm{IBO}}x+n_{\mathrm{u}})+n_{\mathrm{d}}}{\sqrt{P_{\mathrm{sig}}}}, \label{eq:channel}
\end{align}
}

where \(g_{\mathrm{IBO}}\) sets the IBO, \(\arg(\cdot)\) is the phase, \(P_{\mathrm{sig}}\) is the mean noiseless output power over the constellation \(\mathcal X\), and \(z\) is the received sample. The uplink and downlink noises are $n_{\mathrm{u}}\sim\mathcal{CN}(0,g_{\mathrm{IBO}}^2 10^{-\rho_{\mathrm{u}}/10})$ and
\(n_{\mathrm{d}}\sim\mathcal{CN}(0,P_{\mathrm{sig}}10^{-\rho_{\mathrm{d}}/10})\), and \(\rho_{\mathrm{u}}\) and \(\rho_{\mathrm{d}}\) are the uplink and downlink
symbol-energy-to-noise-density ratios \(E_{\mathrm{s}}/N_0\) in decibels, with \(\mathcal{CN}\) the circularly symmetric complex Gaussian distribution. The TWTA uses the Saleh model
\cite{Saleh1981} with AM/AM parameters \(\alpha_{\mathrm a}=2.1587\), \(\beta_{\mathrm a}=1.1517\) and AM/PM parameters \(\alpha_\phi=4.033\), \(\beta_\phi=9.1040\). The demappers operate on the real observation vector
\(\mathbf{z}=[z_{\mathrm I},z_{\mathrm Q}]^\top\in\mathbb R^2\), whose in-phase and quadrature components are \(z_{\mathrm I}=\operatorname{Re}z\) and \(z_{\mathrm Q}=\operatorname{Im}z\).

The links use the DVB second-generation satellite standard for 32-APSK and 16-APSK at rate \(3/4\), with \((\gamma_1,\gamma_2)=(2.84,5.27)\) and \(\gamma=2.85\), respectively
\cite{DeGaudenzi2006,Casini2004,ETSI2014}. The 64-APSK link uses the DVB-S2X \(4+12+20+28\) geometry with code identifier 132/180 and \((\gamma_1,\gamma_2,\gamma_3)=(2.4,4.3,7)\)~\cite{ETSI2024}. Its rate equals the effective short-code rate \(11/15\), but this pairing is not a standard MODCOD. All three links use the short low-density parity-check (LDPC) frame \((N,K)=(16200,11880)\) and the outer Bose--Chaudhuri--Hocquenghem (BCH) code is not applied. Decoding uses normalized min-sum
with scale \(0.75\), at most \(50\) iterations, and syndrome-based early stopping~\cite{ChenFossorier2002}. The fixed IBO is \(3\) dB.

\subsection{Generalized Mutual Information}
Let \(\mathcal X_k^b\) denote symbols whose \(k\)th label bit is \(b\), and let \(\bm{\mu}_x\in\mathbb R^2\) and \(\bm{\Sigma}_x\in\mathbb R^{2\times2}\) be the fitted
per-symbol mean vector and covariance matrix. With equiprobable labels, the cluster-Gaussian bit LLR is
\begin{equation}
L_k(\mathbf{z})=\ln\sum_{x\in\mathcal X_k^0}\mathrm{N}_2(\mathbf{z};\bm{\mu}_x,\bm{\Sigma}_x)
-\ln\sum_{x\in\mathcal X_k^1}\mathrm{N}_2(\mathbf{z};\bm{\mu}_x,\bm{\Sigma}_x),
\label{eq:logmap}
\end{equation}
where \(\mathrm{N}_2(\cdot;\bm{\mu},\bm{\Sigma})\) is the bivariate Gaussian density with mean \(\bm{\mu}\) and covariance \(\bm{\Sigma}\). The centroid MaxLog metric uses a common fitted noise power \(\sigma^2\) and replaces each log-sum by its largest log term:
\begin{equation}
L_k^{\mathrm{MaxLog}}(\mathbf{z})=
\min_{x\in\mathcal X_k^1}\frac{\|\mathbf{z}-\bm{\mu}_x\|^2}{\sigma^2}
-\min_{x\in\mathcal X_k^0}\frac{\|\mathbf{z}-\bm{\mu}_x\|^2}{\sigma^2},
\label{eq:maxlog}
\end{equation}
where \(\sigma^2\) is the fitted total two-dimensional noise power.
For \(\mathbf{b}=(b_1,\ldots,b_m)^\top\), the bitwise GMI is
\begin{equation}
\mathrm{GMI}=m-\sum_{k=1}^{m}\mathbb E\Bigl[
\log_2\Bigl(1+\exp\bigl(-(1-2b_k)L_k\bigr)\Bigr)\Bigr].
\label{eq:gmi}
\end{equation}
The expectation is over transmitted bits and channel observations; GMI serves as the common training and evaluation criterion.

\section{KAN-LUT Demapper}
\subsection{Architecture and Inputs}
The network input is the normalized observation \(\mathbf{z}\) and a bounded downlink-SNR feature. Its two KAN layers are
\begin{align}
h_q&=\sum_{p=1}^{3}\phi_{q,p}(u_p),\qquad \tilde h_q=\tanh(h_q),\label{eq:kanhidden}\\
L_k&=\sum_{q=1}^{H}\psi_{k,q}(\tilde h_q),\label{eq:kan}
\end{align}
where the edge functions use cubic B-splines, with \(i=1,\ldots,G+2\):
\begin{equation}
\begin{aligned}
\phi_{q,p}(u)&=\sum_{i=1}^{G+2}c_{q,p,i}B_i^{(3)}(u),\\
\psi_{k,q}(u)&=\sum_{i=1}^{G+2}d_{k,q,i}B_i^{(3)}(u).
\end{aligned}\label{eq:splines}
\end{equation}
Here \(H\) is the hidden width, \(G\) is the number of uniform grid points on \([-1,1]\), endpoints included, and \(B_i^{(3)}\) are the \(G+2\) clamped cubic B-spline bases. The output layer is also a bank of univariate edges, not a dense linear layer, giving \(mH\) output edges and \(H(3+m)\) edges in total.
The three bounded input features are
\begin{equation}
\begin{aligned}
u_1&=\operatorname{clip}\bigl(z_{\mathrm I}/2.5,-1,1\bigr),\quad
u_2=\operatorname{clip}\bigl(z_{\mathrm Q}/2.5,-1,1\bigr),\\
s&=\operatorname{clip}\bigl((\rho_{\mathrm d}-18)/10,-1,1\bigr),\quad
\mathbf{u}=(u_1,u_2,s)^\top.
\end{aligned}
\label{eq:features}
\end{equation}
The output is the LLR vector \(\mathbf{L}=(L_1,\ldots,L_m)^\top\), with positive entries denoting bit zero.
For 32-APSK, \(m=5\); for 16-APSK, \(m=4\); and for 64-APSK, \(m=6\). Training on \(N_{\mathrm{s}}\) samples minimizes binary cross-entropy with logits \(-L_k\). For the same samples,
\begin{equation}
\mathcal L_{\mathrm{BCE}}=\frac{1}{N_{\mathrm{s}}m}\sum_{n,k}\ln\bigl(1+\mathrm{e}^{-(1-2b_{n,k})L_{n,k}}\bigr),
\label{eq:bce}
\end{equation}
so that \(\mathrm{GMI}=m-(m/\ln 2)\,\mathcal L_{\mathrm{BCE}}\) on the same samples.

\subsection{LUT Evaluation}
Each spline is sampled at 128 uniformly spaced values \(T[i]\); with \(\Delta=(u_{\max}-u_{\min})/127\), the table estimate is
\begin{equation}
\begin{aligned}
\hat\phi(u)&=(1-\lambda)T[i]+\lambda T[i+1],\\
i&=\Bigl\lfloor\frac{u-u_{\min}}{\Delta}\Bigr\rfloor,\qquad
\lambda=\frac{u-u_{\min}}{\Delta}-i,
\end{aligned}
\label{eq:lut}
\end{equation}
with \(i\) clamped to valid adjacent-entry indices. The table therefore requires two reads and one interpolation per edge evaluation.

\begin{proposition}
For an edge \(f\) that is twice continuously differentiable on each table interval intersecting its operating support \(\mathcal U\), linear interpolation on a uniform table of spacing \(\Delta\) satisfies~\cite{deBoor2001}
\[
\sup_{u\in\mathcal U}|f(u)-\hat f(u)|
\leq \frac{\Delta^2}{8}\sup_{u\in\mathcal U}|f''(u)|
=:\varepsilon_f.
\]
For the two-layer network with an intervening \(\tanh\), the resulting LLR error obeys
\[
|\hat L_k-L_k|
\leq \sum_{q=1}^{H}\Bigl[
\varepsilon_{\psi_{k,q}}+
\operatorname{Lip}_{\mathcal V_q}(\psi_{k,q})\sum_{p=1}^{3}\varepsilon_{\phi_{q,p}}
\Bigr].
\]
Here the support is \(\mathcal U=\mathcal U_p\) for \(\phi_{q,p}\) and \(\mathcal U=\mathcal V_q\) for \(\psi_{k,q}\). The set \(\mathcal U_p\) is the union of LUT intervals intersecting the empirical validation range of input feature \(p\), and \(\mathcal V_q\) is the union intersecting the reachable range of \(\tanh(h_q)\), expanded by the hidden-edge interpolation error; \(\operatorname{Lip}_{\mathcal V_q}\) is the Lipschitz constant on \(\mathcal V_q\).
\end{proposition}
\begin{proof}
The edge bound is the linear-interpolation remainder. Hidden-edge errors add and pass through the 1-Lipschitz \(\tanh\); each output edge adds its own interpolation error and a Lipschitz-bounded term from its perturbed input.
\end{proof}
Table~\ref{tab:lutbounds} compares the bound with the maximum error on a support grid and the validation-tail LLR error, for one seed per architecture; every configuration satisfies the bound.
\begin{table}[!t]
\centering
\caption{LUT128 LLR-error bounds and observed validation errors (LLR units, validation-mixture supports).}
\label{tab:lutbounds}
\scriptsize
\setlength{\tabcolsep}{3pt}
\begin{tabular}{lrrr}
\toprule
KAN configuration & Support bound & Support max & Validation $p_{99.9}$ \\
\midrule
H3/G10 & 5.4 & 0.23 & 0.081 \\
H4/G10 & 12 & 0.47 & 0.071 \\
H4/G20 & 57 & 0.53 & 0.24 \\
H8/G10 & 17 & 0.35 & 0.17 \\
H8/G20 & 15 & 0.43 & 0.12 \\
\bottomrule
\end{tabular}

\end{table}

\subsection{ML Training}
MLPs use ReLU with one-layer widths \(8\), \(16\), \(32\), and \(64\) and two-layer widths \(16\times16,32\times32\); KAN widths are \(2,3,4,8\), with \(G\in\{5,10,20\}\).
Training samples downlink \(E_{\mathrm{s}}/N_0\) uniformly over \([14,24]\) dB; the in-phase and quadrature (I/Q) components are divided by \(2.5\) and clipped to \([-1,1]\), and the SNR feature uses
\eqref{eq:features}. Adam uses gradient clipping and cosine learning-rate decay. For 32-APSK, each architecture receives 60 hyperparameter optimization (HPO) trials, evenly distributed across
the three intervals \([10^{-4},10^{-2}]\), \([3\times10^{-3},10^{-1}]\), and \([3\times10^{-2},3\times10^{-1}]\), on \(10^7\) symbols; the smallest rate within
0.005 bit of best validation GMI is selected. The \(\{10^7,3\times10^7,10^8\}\)-symbol ladder selects the first rung whose next-rung gain is below
0.005 bit, or the largest rung if none qualifies. Each model is trained with five independent seeds. For 16- and 64-APSK, four models use
20 log-uniform trials over \([0.03,0.3]\) on \(10^7\) symbols, then the same ladder.
At LUT128, H2 reaches at most 3.758 bit/symbol mean validation GMI (H2/G5), versus 3.965 bit/symbol for H3/G10 (five seeds each), and is therefore not included in the FER evaluation.

\subsection{Complexity Analysis}
Let \(\mathbf{N}\) collect the per-symbol multiply, add, compare, transcendental, and LUT-read counts, and let \(\mathbf{w}\) hold the corresponding weights. The weighted cost is
\begin{equation}
C_{\mathrm W}=\mathbf{w}^\top\mathbf{N},
\label{eq:cost}
\end{equation}
with \(\mathrm{W1}\) weights \((1,1,1,10,1)\) and \(\mathrm{W2}\) weights \((1,0.5,0.5,20,2)\). Table~\ref{tab:costs} reports operation counts, LUT reads, memory, and weighted costs.
\begin{table}[!t]
\centering
\caption{Per-symbol operation counts (M/A/C/E/L), memory, and W1/W2 costs. Bold: KAN-LUT; italics: LogMAP-CG reference.}
\label{tab:costs}

\footnotesize
\setlength{\tabcolsep}{2pt}
\begin{tabular}{lcrrr}
\toprule
Method & Ops (M/A/C/E/L) & Mem KiB & W1 & W2 \\
\midrule
\multicolumn{5}{l}{16-APSK (four LLR outputs)} \\
\midrule
\textit{LogMAP-CG shared} & \textit{(96/172/15/24/0)} & \textit{0.438} & \textit{523} & \textit{669.5} \\
MaxLog & (49/52/56/0/0) & 0.129 & 157 & 103.0 \\
MLP-64 & (451/449/68/0/0) & 2.02 & 968 & 709.5 \\
MLP-32$\times$32 & (1251/1249/68/0/0) & 5.14 & 2568 & 1909.5 \\
\textbf{KAN-H4-LUT G20} & \textbf{(38/84/18/4/56)} & \textbf{14.0} & \textbf{236} & \textbf{281.0} \\
\textbf{KAN-H8-LUT G20} & \textbf{(70/168/26/8/112)} & \textbf{28.0} & \textbf{456} & \textbf{551.0} \\
\midrule
\multicolumn{5}{l}{32-APSK (five LLR outputs)} \\
\midrule
\textit{LogMAP-CG shared} & \textit{(192/379/31/42/0)} & \textit{0.875} & \textit{1022} & \textit{1237.0} \\
MaxLog & (97/101/150/0/0) & 0.254 & 348 & 222.5 \\
Ring-sector & (77/81/117/1/0) & 0.684 & 285 & 196.0 \\
MLP-8 & (67/65/12/0/0) & 0.301 & 144 & 105.5 \\
MLP-16 & (131/129/20/0/0) & 0.582 & 280 & 205.5 \\
MLP-32 & (259/257/36/0/0) & 1.14 & 552 & 405.5 \\
MLP-64 & (515/513/68/0/0) & 2.27 & 1096 & 805.5 \\
MLP-16$\times$16 & (387/385/36/0/0) & 1.64 & 808 & 597.5 \\
MLP-32$\times$32 & (1283/1281/68/0/0) & 5.27 & 2632 & 1957.5 \\
\textbf{KAN-H3-LUT G10} & \textbf{(33/71/16/3/48)} & \textbf{12.0} & \textbf{198} & \textbf{232.5} \\
\textbf{KAN-H4-LUT G20} & \textbf{(42/95/18/4/64)} & \textbf{16.0} & \textbf{259} & \textbf{306.5} \\
\textbf{KAN-H8-LUT G10} & \textbf{(78/191/26/8/128)} & \textbf{32.0} & \textbf{503} & \textbf{602.5} \\
\midrule
\multicolumn{5}{l}{64-APSK (six LLR outputs)} \\
\midrule
\textit{LogMAP-CG shared} & \textit{(384/826/63/76/0)} & \textit{1.75} & \textit{2033} & \textit{2348.5} \\
MaxLog & (193/198/372/0/0) & 0.504 & 763 & 478.0 \\
MLP-64 & (579/577/68/0/0) & 2.52 & 1224 & 901.5 \\
MLP-32$\times$32 & (1315/1313/68/0/0) & 5.40 & 2696 & 2005.5 \\
\textbf{KAN-H4-LUT G20} & \textbf{(46/106/18/4/72)} & \textbf{18.0} & \textbf{282} & \textbf{332.0} \\
\textbf{KAN-H8-LUT G20} & \textbf{(86/214/26/8/144)} & \textbf{36.0} & \textbf{550} & \textbf{654.0} \\
\bottomrule
\end{tabular}

\end{table}
With 32-bit entries and 128 points per input dimension, a direct three-input LUT for five LLRs needs
\(128^3\times5\times4~\mathrm{B}=40.0~\mathrm{MiB}\). An edge table uses
\(128\times4~\mathrm{B}=512~\mathrm{B}\); H8 has 64 edges and therefore uses
32~KiB. Edge-operation counts are independent of spline grid size \(G\), while storage grows with \(G\). Classical memory holds the fitted means, covariances, or centroids.

Let \(d=3\), \(H\) be the KAN hidden width, \(m\) the output count, and \(N_{\mathrm e}=H(d+m)\) the total number of edge functions. For one symbol, the KAN-LUT counts, including input scaling and clipping, are
\[
\begin{aligned}
\mathbf{N}_{\mathrm{KAN}}=\bigl(&N_{\mathrm e}+d+H+3,\;3N_{\mathrm e}+d-m+1,\\
&2(d+H)+4,\;H,\;2N_{\mathrm e}\bigr),
\end{aligned}
\]
with entries ordered as in \(\mathbf{N}\). Thus, there are \(2N_{\mathrm e}\) table reads and \(N_{\mathrm e}\) interpolation multiplications; the \(H\) intervening \(\operatorname{tanh}\)
evaluations are counted as transcendental operations. For an MLP with layer widths \((n_0,\ldots,n_{r+1})=(d,H_1,\ldots,H_r,m)\), the weight count is \(n_{\mathrm W}=\sum_{j=1}^{r+1}n_{j-1}n_j\). Its counts are
\((n_{\mathrm W}+3,n_{\mathrm W}+1,\sum_{j=1}^{r}H_j+4,0,0)\), including the shared input-feature operations. For balanced bit labels over \(M\) constellation points, the cluster-Gaussian
LogMAP-CG shared-exponential counts are
\[
\begin{aligned}
\mathbf{N}_{\mathrm{LogMAP,shared}}=\bigl(&6M,\;7M+m(M-1),\\
&M-1,\;M+2m,\;0\bigr),
\end{aligned}
\]
where \(n_{k,b}=|\mathcal X_k^b|\) and \(\sum_{k=1}^{m}[(n_{k,0}-1)+(n_{k,1}-1)+1]=m(M-1)\). A shared maximum stabilizes symbol likelihoods and allows exponentials to be reused across bit subsets. A per-bit log-sum-exp evaluation gives the same LLRs at a higher cost, so all LogMAP-CG costs use the shared form. Centroid MaxLog uses
\((3M+1,3M+m,m(M-2),0,0)\).

\section{Numerical Results}\label{sec:results}

For all three constellations, the uplink \(E_{\mathrm{s}}/N_0\) is set 1 dB above the lowest tested value at which LogMAP-CG showed a clear FER waterfall, giving 17~dB and 13~dB for 32- and 16-APSK and 19~dB for 64-APSK. Validation data are used for HPO, training-budget selection, and operating-point selection while test frames are held out. Required \(E_{\mathrm{s}}/N_0\) at FER \(10^{-3}\) used model-specific \(0.5\)-dB scout and \(0.1\)-dB dense bracketing, at least 100 errors per point where permitted, and a 300{,}000-frame cap. We interpolated \(\log_{10}(\mathrm{FER})\) against \(E_{\mathrm{s}}/N_0\); crossings beyond the upper boundary are censored. For each seed, all neural models are evaluated on the same test frames. All demappers use the same decoder and channel-draw rule within each constellation. Inference uses paired two-sided \(t\)-tests with 95\% CIs.

Table~\ref{tab:esn0} reports required \(E_{\mathrm{s}}/N_0\), sample standard deviations, and gaps to LogMAP-CG and the best MLP. \FGR{fig:pareto} shows the \(\mathrm{W1}\) trade-off, and the \(\mathrm{W2}\) costs in Table~\ref{tab:costs} preserve the same ordering. The KAN-H8-LUT G10 requires 18.15~dB, which is 0.09~dB above LogMAP-CG and 1.02~dB below MaxLog, at 49\%/49\% of the shared LogMAP-CG cost under W1/W2.
\begin{figure}[!t]
\centering
\includegraphics[width=\columnwidth]{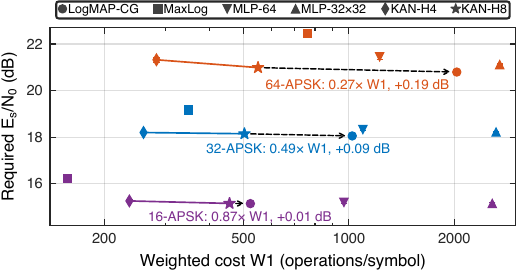}
\setlength{\abovecaptionskip}{0pt}
\caption{Required \(E_{\mathrm{s}}/N_0\) at FER \(10^{-3}\) versus W1 cost for 16-, 32-, and 64-APSK. Markers show seed means and bars sample standard deviations. KAN-H8 uses G10 at 32-APSK and G20 otherwise. The orders use different operating points, so only gaps within an order are comparable.}

\label{fig:pareto}
\end{figure}

\begin{table}[!t]
\centering
\caption{Mean required \(E_{\mathrm{s}}/N_0\) (dB), SD, and difference \(\Delta\) (method minus reference) to LogMAP-CG and the best MLP. H8 uses G10 (32-APSK) and G20 (16-APSK). Bold: best learned mean and gaps within 0.1~dB of LogMAP-CG.}
\label{tab:esn0}
\setlength{\tabcolsep}{2pt}
\scriptsize
\renewcommand{\arraystretch}{0.82}
\resizebox{\columnwidth}{!}{%
\begin{tabular}{lrrrrrrrr}
\toprule
Method & \multicolumn{4}{c}{32-APSK} & \multicolumn{4}{c}{16-APSK} \\
\cmidrule(lr){2-5}\cmidrule(lr){6-9}
 & Mean & SD & $\Delta$CG & $\Delta$MLP & Mean & SD & $\Delta$CG & $\Delta$MLP \\
\midrule
LogMAP-CG & $18.05$ & \textemdash & $0.00$ & $-0.17$ & $15.15$ & \textemdash & $0.00$ & $-0.01$ \\
MaxLog & $19.17$ & \textemdash & $1.12$ & $0.95$ & $16.22$ & \textemdash & $1.07$ & $1.06$ \\
Ring-sector & $19.17$ & \textemdash & $1.12$ & $0.95$ & \textemdash & \textemdash & \textemdash & \textemdash \\
MLP-8 & $>22.00$ & \textemdash & \textemdash & \textemdash & \textemdash & \textemdash & \textemdash & \textemdash \\
MLP-16 & $19.75$ & $1.01$ & $1.69$ & $1.53$ & \textemdash & \textemdash & \textemdash & \textemdash \\
MLP-32 & $18.47$ & $0.05$ & $0.42$ & $0.25$ & \textemdash & \textemdash & \textemdash & \textemdash \\
MLP-64 & $18.33$ & $0.03$ & $0.27$ & $0.11$ & $15.20$ & $0.04$ & \textbf{\boldmath$0.05$} & $0.04$ \\
MLP-32$\times$32 & $18.22$ & $0.04$ & $0.17$ & $0.00$ & \textbf{\boldmath$15.16$} & $0.03$ & \textbf{\boldmath$0.01$} & $0.00$ \\
MLP-16$\times$16 & $18.79$ & $0.39$ & $0.74$ & $0.57$ & \textemdash & \textemdash & \textemdash & \textemdash \\
KAN-H3-LUT G10 & $18.58$ & $0.29$ & $0.53$ & $0.36$ & \textemdash & \textemdash & \textemdash & \textemdash \\
KAN-H4-LUT G20 & $18.20$ & $0.04$ & $0.15$ & $-0.02$ & $15.26$ & $0.04$ & $0.12$ & $0.11$ \\
KAN-H8-LUT & \textbf{\boldmath$18.15$} & $0.03$ & \textbf{\boldmath$0.09$} & $-0.07$ & \textbf{\boldmath$15.16$} & $0.03$ & \textbf{\boldmath$0.01$} & $0.00$ \\
\bottomrule
\end{tabular}
}

\end{table}

Table~\ref{tab:tests} lists the paired KAN-minus-MLP gaps, where negative values favor the KAN. For 32-APSK, both H8 grids require 0.07~dB less than the best MLP, MLP-32$\times$32, a small but statistically significant difference. The smaller H4 model is not significantly different from MLP-32$\times$32 at either grid. For 16-APSK, H8/G20 matches both MLPs, whereas H4/G20 is slightly but significantly worse. A KAN therefore needs the wider H8 hidden layer to match or exceed the MLPs.

\begin{table}[!t]
\centering
\caption{Paired KAN-minus-MLP gaps (dB), 95\% CIs, and \(p\)-values; 16- and 64-APSK use G20. Bold/italics: significant (\(p<0.05\)) in favor of the KAN/MLP.}
\label{tab:tests}

\scriptsize
\setlength{\tabcolsep}{2pt}
\resizebox{\columnwidth}{!}{%
\begin{tabular}{lrrr}
\toprule
Comparison & Mean gap (dB) & 95\% CI (dB) & $p$ \\
\midrule
\multicolumn{4}{l}{16-APSK} \\
\midrule
\textit{KAN-H4-LUT G20 vs MLP-32$\times$32} & \textit{\ensuremath{\mathit{0.11}}} & \textit{\ensuremath{\mathit{[0.02, 0.19]}}} & \textit{\ensuremath{\mathit{0.027}}} \\
\textit{KAN-H4-LUT G20 vs MLP-64} & \textit{\ensuremath{\mathit{0.06}}} & \textit{\ensuremath{\mathit{[0.04, 0.09]}}} & \textit{\ensuremath{\mathit{0.0022}}} \\
KAN-H8-LUT G20 vs MLP-32$\times$32 & $0.00$ & $[-0.06, 0.06]$ & $>0.99$ \\
KAN-H8-LUT G20 vs MLP-64 & $-0.04$ & $[-0.12, 0.03]$ & $0.17$ \\
\midrule
\multicolumn{4}{l}{32-APSK} \\
\midrule
KAN-H4-LUT G10 vs MLP-32$\times$32 & $0.01$ & $[-0.06, 0.09]$ & $0.71$ \\
KAN-H4-LUT G20 vs MLP-32$\times$32 & $-0.02$ & $[-0.09, 0.05]$ & $0.45$ \\
\textbf{KAN-H8-LUT G10 vs MLP-32$\times$32} & \textbf{\boldmath$-0.07$} & \textbf{\boldmath$[-0.12, -0.03]$} & \textbf{\boldmath$0.011$} \\
\textbf{KAN-H8-LUT G20 vs MLP-32$\times$32} & \textbf{\boldmath$-0.07$} & \textbf{\boldmath$[-0.11, -0.04]$} & \textbf{\boldmath$0.0037$} \\
\midrule
\multicolumn{4}{l}{64-APSK} \\
\midrule
\textit{KAN-H4-LUT G20 vs MLP-32$\times$32} & \textit{\ensuremath{\mathit{0.21}}} & \textit{\ensuremath{\mathit{[0.11, 0.31]}}} & \textit{\ensuremath{\mathit{0.0044}}} \\
KAN-H4-LUT G20 vs MLP-64 & $-0.13$ & $[-0.37, 0.11]$ & $0.21$ \\
\textbf{KAN-H8-LUT G20 vs MLP-32$\times$32} & \textbf{\boldmath$-0.13$} & \textbf{\boldmath$[-0.17, -0.08]$} & \textbf{\boldmath$0.0013$} \\
\textbf{KAN-H8-LUT G20 vs MLP-64} & \textbf{\boldmath$-0.46$} & \textbf{\boldmath$[-0.61, -0.31]$} & \textbf{\boldmath$0.0010$} \\
\bottomrule
\end{tabular}
}

\end{table}

\FGR{fig:fer} shows representative 32-APSK FER curves.

\begin{figure}[!t]
\centering
\includegraphics[width=\columnwidth]{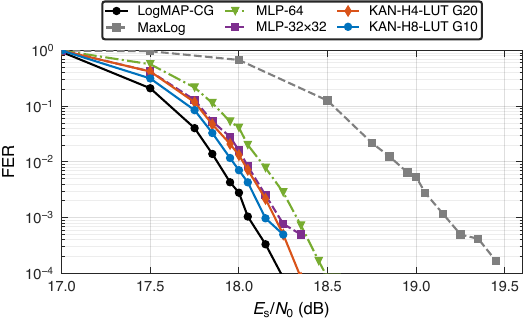}
\setlength{\abovecaptionskip}{0pt}
\caption{Representative 32-APSK FER curves; neural models use median-required-\(E_{\mathrm{s}}/N_0\) seeds; zero-error points are omitted.}
\label{fig:fer}
\end{figure}

\begin{figure}[!t]
\centering
\includegraphics[width=\columnwidth]{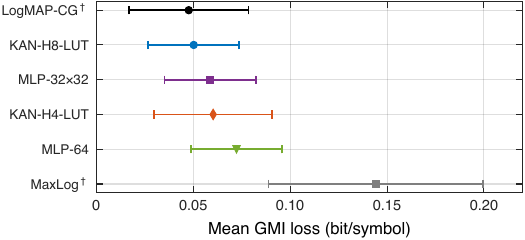}
\setlength{\abovecaptionskip}{0pt}
\caption{Mean GMI loss to matched (per-point refitted) LogMAP-CG over the 27 operating points; bars show sample standard deviation across points, not uncertainty. KAN rows are G20. \(\dagger\) Classical demappers fitted at the training point only.}
\label{fig:mismatch}
\end{figure}

To test out-of-sample performance, we evaluate the demappers away from their training point (IBO of 3~dB, uplink \(E_{\mathrm{s}}/N_0\) of 17~dB). The grid spans IBO \(\{2,3,4\}\)~dB, uplink \(E_{\mathrm{s}}/N_0\) \(\{15,17,20\}\)~dB, and downlink \(E_{\mathrm{s}}/N_0\) \(\{17,18,19\}\)~dB, giving 27 points with 200{,}000 common channel draws each. At every point, the reference is LogMAP-CG refitted to that point, and each demapper is scored by its GMI loss relative to this reference. The neural models are used as trained, and the classical LogMAP-CG and MaxLog keep their training-point fits. This study uses H4/G20 and H8/G20. Fig.~\ref{fig:mismatch} shows the mean loss over the 27 points.

At each point, we compare each KAN with each MLP using a paired test and apply Holm correction within each of the four KAN--MLP families. KAN-H8-LUT is significantly better than MLP-32$\times$32 at 7 points and better than MLP-64 at 26 points, and it is not significantly worse than either MLP at any point. KAN-H4-LUT gives mixed results: against MLP-32$\times$32 it is better at 4 points and worse at 2, and against MLP-64 it is better at 16 points and worse at 1. These differences are small, below 0.01~bit/symbol on average.
Averaged over the 27 points, KAN-H8-LUT loses 0.050~bit/symbol relative to the matched LogMAP-CG, less than MLP-32$\times$32 (0.059), KAN-H4-LUT (0.060), and MLP-64 (0.072). This is close to the 0.048~bit/symbol lost by LogMAP-CG fitted only at the training point, while MaxLog with a training-point fit loses 0.144~bit/symbol. The KAN thus tolerates operating-point mismatch about as well as the model-based demapper it approximates.

Table~\ref{tab:lutablation} shows how LUT size trades memory for accuracy. For \(G\leq10\), LUT64 halves the memory of LUT128 with negligible GMI loss. At \(G=20\), LUT64 loses up to 4.7 and 2.5~mbit/symbol for H4 and H8, compared with 0.29 and 0.11~mbit/symbol at LUT128. This agrees with the \(\Delta^2\sup|f''|/8\) term of the interpolation bound, because a finer spline grid allows edges of higher curvature, which a coarse table follows less accurately.

\FGR{fig:pareto} also shows the cost--accuracy trade-off for 64-APSK. From 16- to 32- to 64-APSK, the W1 cost of LogMAP-CG roughly doubles at each step, from 523 to 1022 to 2033, because its operation count grows with the number of constellation points. The cost of KAN-H8-LUT rises only from 456 to 503 to 550, because only the output edges depend on the constellation. The KAN cost relative to LogMAP-CG therefore falls from 87\% to 49\% to 27\%. For 64-APSK, KAN-H8-LUT G20 is 0.19~dB above LogMAP-CG and 1.47~dB below MaxLog. As Table~\ref{tab:tests} shows, it requires 0.13~dB less than MLP-32$\times$32 and 0.46~dB less than MLP-64, and both differences are statistically significant. The gap to LogMAP-CG therefore widens slightly with order, while the cost advantage grows much faster. The H8 tables need 36~KiB, against 48~MiB for a direct LUT.

Training on \(3\times10^{8}\) instead of \(10^{8}\) symbols still raised validation GMI by up to 0.015~bit/symbol for both families, so the small FER gaps could shift. Gaps of about 0.1~dB are small in link-budget terms, so the main benefit of KAN-LUT is its lower cost.

\begin{table}[!t]
\centering
\caption{Spline-minus-LUT GMI loss (mbit/symbol, median/maximum over five runs) and LUT memory (KiB); bold: LUT128.}
\label{tab:lutablation}
\footnotesize
\setlength{\tabcolsep}{2pt}
\scriptsize
\begin{tabular}{lrrrr}
\toprule
KAN Grid & LUT16 & LUT32 & LUT64 & \textbf{LUT128} \\
\midrule
H4/G5 & 37/44 & 2/2.5 & $<$0.1/$<$0.1 & \textbf{\boldmath$<$0.1/$<$0.1} \\
H4/G10 & 200/260 & 12/25 & 0.64/1.3 & \textbf{\boldmath$<$0.1/0.12} \\
H4/G20 & 783/1292 & 64/117 & 3.2/4.7 & \textbf{0.12/0.29} \\
H8/G5 & 37/51 & 1.7/2.8 & $<$0.1/0.12 & \textbf{\boldmath$<$0.1/$<$0.1} \\
H8/G10 & 84/90 & 4.5/7.5 & 0.24/0.41 & \textbf{\boldmath$<$0.1/$<$0.1} \\
H8/G20 & 158/697 & 9.2/46 & 0.65/2.5 & \textbf{\boldmath$<$0.1/0.11} \\
\midrule
Mem. H4 KiB & 2 & 4 & 8 & \textbf{16} \\
Mem. H8 KiB & 4 & 8 & 16 & \textbf{32} \\
\bottomrule
\end{tabular}
\end{table}

\section{Conclusion}

This paper introduces a KAN-LUT soft demapper for APSK over a nonlinear satellite channel, in which the edges of a two-layer KAN are stored as 128-entry LUTs. We bound the table error and compare the demapper with MLPs and with per-cluster LogMAP-CG and MaxLog demappers. For 16-, 32-, and 64-APSK, KAN-LUT stays within 0.2~dB of LogMAP-CG, and its relative cost falls from 87\% to 27\% as the order grows. It matches or slightly exceeds the best MLP at about one fifth of its cost, remains competitive under operating-point mismatch, and needs tens of kilobytes of tables instead of tens of megabytes. KAN-LUT is thus a lower-cost approximation to log-MAP demapping rather than a more accurate one. Future work includes fixed-point tables, hardware measurements, channels with memory, and multiple back-off levels.

\section*{Acknowledgment}
This work was supported by the Natural Sciences and Engineering Research Council (NSERC) of Canada Alliance grant ALLRP 579869-22 (Artificial Intelligence Enabled Harmonious Wireless Coexistence for 3D Networks [3D-HARMONY]).

\bibliographystyle{IEEEtran}
\bibliography{main}
\balance

\end{document}